\documentclass[a4paper,twocolumn]{emulateapj}
\usepackage{amstext}
\usepackage{apjfonts}
\usepackage{amsmath}
\usepackage{amssymb}

\newcommand{\nar}{NaR}

\newcommand{\RNum}[1]{\uppercase\expandafter{\romannumeral #1\relax}}

\usepackage[colorlinks=true,linkcolor=blue,citecolor=blue]{hyperref}
\begin{document}


\title{THE ROLE OF NUCLEAR STAR CLUSTERS IN ENHANCING SUPERMASSIVE BLACK HOLE FEEDING RATES DURING GALAXY MERGERS}

\author{J. P. Naiman\altaffilmark{1}, E. Ramirez-Ruiz\altaffilmark{2}, J. Debuhr\altaffilmark{3}, C.-P. Ma\altaffilmark{4}}
\altaffiltext{1}{Harvard-Smithsonian Center for Astrophysics, Cambridge, MA 02138}
\altaffiltext{2}{Department of Astronomy and 
  Astrophysics, University of California, Santa Cruz, CA
  95064}
\altaffiltext{3}{Center for Research in Extreme Scale Technologies, Indiana University, Bloomington IN, 47404}
 \altaffiltext{4}{Department of Astronomy, University of California, Berkeley, CA, 94720}
   
 \begin{abstract} 
 During galaxy mergers the gas falls to the center, triggers star formation, and feeds the rapid growth of supermassive black holes (SMBHs). SMBHs respond to this fueling by supplying energy back to the ambient gas. Numerical studies suggest that this feedback is necessary to explain why the properties of SMBHs and the formation of bulges are closely related. This intimate link between the SMBH's mass and the large scale dynamics and luminosity of the host has proven to be a difficult issue to tackle with simulations due to the inability to resolve all the relevant length scales simultaneously. In this paper we simulate SMBH growth at high-resolution with {\it FLASH}, accounting for the gravitational focusing effects of nuclear star clusters (NSCs), which appear to be ubiquitous in galactic nuclei.   In the simulations, the NSC core is resolved  by a minimum  cell size of about  0.001 pc or approximately $10^{-3}$ of the cluster's radius. We discuss the conditions required for effective gas funneling to occur, which are mainly  dominated by a relationship between NSC velocity dispersion and the local sound speed, and provide a sub-grid prescription for the augmentation of central SMBH accretion rates in the presence of NSCs. For the conditions expected to persist in the centers of merging galaxies, the resultant large central gas densities in NSCs should produce drastically enhanced embedded SMBH accretion rates - up to an order of magnitude increase can be achieved for gas properties resembling those in large-scale galaxy merger simulations. This will naturally result in faster black hole growth rates and higher luminosities than predicted by the commonly used Bondi-Hoyle-Lyttleton accretion formalism.
\end{abstract}


\section{Introduction}

Supermassive black holes (SMBHs; $M \gtrsim 10^5 \, M_\odot$) are  inferred to reside in most galactic nuclei \citep{kormendy1995, magorrian1998,laor2000,ferra2006,shan2009}.   
When efficiently supplied with gas, these objects can produce some of the most luminous sources in the Universe \citep{sil2008,hue2010,falocco2012,koss2012}.   
Recent panchromatic surveys have shown  that close dual AGN are comparatively more  luminous at high energies  than their isolated counterparts, suggesting that the merger process is intricately tied to the feeding history of the waltzing SMBHs \citep{gul2009a,gul2009b,koss2012,liu2013}.
 
Gas supply to central SMBHs during galaxy mergers directly affects the growth and luminosity of these objects.   Therefore, understanding how mass is bestowed to SMBHs  results in predictions of their number density and luminosity distribution \citep{silk1998,komossa2003}. 
In cosmological simulations of merging galaxies, some form of the classical Bondi-Hoyle-Lyttleton (BHL) accretion prescription is usually implemented to estimate SMBH feeding rates \citep{springel2005,kuro2009,fabjan2010,li2012,jeon2012,choi2012,choi2013,hirschmann2013,newton2013,blecha2013,angles2013,barai2014,gabor2014}.  The use of this recipe assumes that the properties of gas at 50-100~pc (generally determined by the resolution scale length) accurately models the mass accretion rate onto the SMBH
\citep{johansson2009}. 

A significant fraction of SMBHs are expected to be embedded in nuclear star clusters \citep[NSCs;][]{boker2002,boker2004,walcher2005,graham2009,boker2010,seth2010}.  Because these  NSCs are typically more massive than the central SMBH, they can significantly alter the gas flow at scales which are commonly  unresolved in cosmological simulations \citep[$\approx 1-5 \, {\rm pc}$;][]{naiman2011}. As a result, NSCs could   provide
an efficient mechanism for funneling gas towards the central black hole. Accurately determining the mass accretion history of SMBHs 
has important  consequences not only for their growth history, but also for the evolution of the host galaxy \citep{king2003,croton2006,hopkins2007,sij2007,dimatteo2008,primack2008,booth2009,debuhr2011,debuhr2012}. 
For this reason, pinning down the dominant mechanism by which gas at large scales is funneled into the SMBH is essential for understanding their role in mediating galaxy evolution.

In this paper,  we make use of simulations to investigate how the accretion rate of SMBHs might be  enhanced during galaxy mergers when they are embedded in massive and compact NSCs.   
Hydrodynamical models with multiple levels of refinement are used  to capture both the large scale gas flow around the SMBH and NSC complex as well as the small scale accretion onto the central sink.  We model the SMBH and the NSC as static potentials and simulate the structure of supersonic gas flows within their  combined gravitational field  in order to quantify the gas accretion rate in the inner tenths of parsecs. Simulations are performed  for adiabatic and isothermal flows as well as  for  wide range of NSC  radii and  ambient  conditions.
These well-resolved, three dimensional simulations are used to generate sub-grid accretion prescriptions for larger scale simulations, thus providing a more accurate estimate of the feeding rates of SMBHs in cosmological simulations as well as a better determination of the expected luminosities 
of dual AGNs.  We test the effects of this accretion rate enhancement on the growth of SMBHs in the galaxy merger simulations of \cite{debuhr2011,debuhr2012} and discuss what combination of gas and gravitational potential parameters results in a significant augmentation to the mass accretion rate.

This paper is organized as follows. In Section \ref{section:anal} we review the commonly used sub-grid prescriptions for calculating  the mass accretion rates onto central massive black holes during galaxy merger simulations and suggest a modification  in the presence of a NSC.  A numerical   scheme  aimed at calculating   black hole growth at high resolution, accounting for the gravitational focusing effects of NSC, is presented in Section \ref{section:methods}. The conditions necessary for NSCs to collect ambient gas and, in turn, enhance the mass accretion rate of  SMBHs are derived  in Section \ref{sec:cond}.  In Section \ref{section:mergerMods} we use the central gas densities from realistic SPH cosmological merger simulations to estimate the augmentation to 
SMBHs mass accretion rates in the presence of a NSC as a function of  time. We summarize our findings in Section \ref{section:summary}.

\section{Accretion Flows  Modified by the Presence of a NSC} \label{section:anal}

SMBHs can trigger  nuclear activity only as long as they interact with the surrounding gas.
The way in which gas flows into a SMBH depends largely on the conditions where material is injected. 
The mass accretion rate into SMBHs  in cosmological simulations is commonly estimated using analytical prescriptions based on the large scale gas  structures in the centers of the simulated merging galaxies.  The BHL formalism assumes the gas is accreted spherically symmetrically \citep{johansson2009}: 
\begin{equation}
\dot{M}_{\rm bondi} = 4 \pi (G M)^2 \rho_\infty (v^2 + c_{\rm s}^2)^{-3/2}, 
\end{equation}
where $v$ is the relative velocity of the object through the external gas whose sound speed and 
density  are given by $c_{\rm s}$ and $\rho_\infty$, respectively \citep{edgar2004}.  

The flow pattern is dramatically altered if the inflowing gas has a small amount of angular momentum. If the inflowing gas is injected more or less isotropically from large $r$, but has specific angular momentum per unit mass such that $l^2/GM \gg r_{\rm g}$, then the quasi-spherical approximation will break  down and the gas will have sufficient angular momentum  to orbit the SMBH (here $r_{\rm g}$ is the SMBH's Schwarzschild radius).  Viscous torques will then cause the gas to sink into the equatorial plane of the SMBH.  In recent works, such departures from spherical symmetry have been accounted 
for by assuming the accretion proceeds through a disk: 
\begin{equation}
\dot{M}_\nu = 3 \pi \alpha \Sigma c_{\rm s}^2 \Omega^{-1}
\label{eq:cpmmacc}
\end{equation}
where $\Sigma$ is the average surface gas density of the accretion disk, $\Omega$ is the average rotational angular frequency, and $\alpha$ is a dimensionless parameter dictating the strength of turbulent viscosity in the disk
\citep{debuhr2011,debuhr2012}.  Both prescriptions rely on an understanding of  the gas properties at   $r \gtrsim$100~pc in order  to determine the mass accretion rate onto the central SMBH.

As discussed in \cite{naiman2009,naiman2011}, a NSC moving at 
a low Mach number  through relatively cold gas 
 can drastically increase the gas density in its interior with respect to that of the external medium. 
 Analytical estimates suggest that in the central regions of a NSC the expected density enhancement is given by
\begin{equation}
{\rho(r = 0) \over  \rho_\infty} \approx  \left[ 1 + {G M_{\rm c} (\gamma-1) \over  c_\infty^{2} r_{\rm c}} \right]^{1/(\gamma-1)}, 
\label{eq:rhostationary}
\end{equation}
where the core radius, $r_{\rm c}$, is related to the cluster velocity dispersion by $\sigma_{V}^2 \approx GM_{\rm c}/r_{\rm c}$.  The gas properties characterized  here by $\rho_\infty$ and $c_\infty$ are the density and sound speed at   $r\gg r_{\rm c}$.
Here $\gamma$ denotes the adiabatic index of the flow and $M_{\rm c}$ designates  the mass of the NSC \citep{naiman2011}.    Given this density  enhancement  in the core of the NSC, the accretion rate of the  embedded SMBH would be amplified by a factor of $\dot{M}_{\rm bh+c} \propto \rho(r=0)/c_{\rm s}^3(r=0)$ where $c_{\rm s} (r=0)=c_{\rm s,nsc}$  is the sound speed of the gas within the NSC and we have assumed here that  accretion proceeds at the classical  Bondi rate \citep{bondi1952}.

Even a small amount of angular momentum can make a big difference, breaking the spherical symmetry of the inflowing gas
and yielding  an accretion  disk instead of the radial flow assumed  in equation \ref{eq:rhostationary}.  If we instead suppose  that accretion onto the central SMBH  proceeds by viscous torques and  require   the average surface density of the disk to be proportional to  $\rho_{\rm nsc}$,  the   mass accretion rate can be approximated as \citep{debuhr2011}
\begin{equation}
\dot{M}_{\rm bh+c} = 3 \pi \alpha \Sigma_{\rm nsc} c_{\rm s, nsc}^2 \Omega^{-1}
\label{eq:mdot}
\end{equation}
with $c_{\rm s, nsc} = K \gamma \rho_{\rm nsc} ^{\gamma-1}$, and 
\begin{equation}
{\Sigma_{\rm nsc} \over \Sigma} = {\rho_{\rm nsc} \over  \rho_\infty } = \max\left\{\left( 1 + \frac{G M_{\rm c} [\gamma -1 ]}{c_\infty^2 r_{\rm c} [ 1 + \mu_\infty^2]} \right)^{\frac{1}{\gamma-1}}
\left( \frac{\mu_\infty^2}{\mu_\infty^2 + 1}\right)^{\frac{3}{2}},1\right\}.
\label{eq:rho}
\end{equation}
Here $\rho_{\rm nsc}$
is the modified density in the NSC's interior and $\mu_\infty = v_\infty/c_\infty$ is the Mach number of the SMBH+NSC complex  with respect to the ambient  gas. The resultant accretion rate would, in this limit,  be amplified by a factor $\dot{M}_{{\rm bh+c}} \propto \rho(r=0) c_{\rm s}^2(r=0)$.

Contrary to the classical BHL case, the motion of the SMBH+NSC complex with respect to the ambient medium  does not  result  in a considerable change in the rate of mass accretion onto the SMBH. This is because  equation \ref{eq:rho} accounts for the protection provided by the large scale  NSC which forms  a quasi-hydrostatic envelope  around  the SMBH.  Having said this, the strength of this protection is  indeed slightly diminished  because the capture radius  decreases  due to the relative motion of the potential  \citep{naiman2011}. As a result, the expected enhancement in the  gas density of the NSC core is  indeed moderately  smaller  than the one given by the stationary formula (equation \ref{eq:rhostationary}). More importantly, we have assumed here that  the influence of the NSC dominates  the gravitational potential, which implies $M_{\rm c} \gtrsim M_{\rm bh}$.  When  $M_{\rm c} \lesssim M_{\rm bh}$, the formation of a quasi-hydrostatic envelope is  inhibited.  The presence of a compact and massive NSC with  $M_{\rm c} \gtrsim M_{\rm bh}$ is thus required in order to  significantly  increase the accretion rate onto the  central SMBHs during a galaxy merger.  The degree by which the accretion rate is enhanced by the presence  of a NSC is 
 studied  with numerical simulations in the remaining of the paper.

\section{Simulating Accretion onto SMBH\lowercase{s} embedded in NSC\lowercase{s}} \label{section:methods}

To examine the ability of  a NSC to collect ambient gas  and, in turn, enhance the mass accretion rate onto the central SMBH,
we simulate the  SMBH~+~NSC complex as a  gravitational potential  $\Phi = \Phi_{\rm c} + \Phi_{\rm bh}$ moving through ambient gas with FLASH, a parallel, adaptive mesh
refinement hydrodynamics code \citep{fryxell}.  A smooth  potential, given by 
\begin{equation}
\Phi_{\rm c} = {G M_{\rm c} \over (r^2 + r_{\rm c}^2)^{1/2}}, 
\end{equation}
provides an accurate description of the NSC potential  given the cluster mass, $M_{\rm c}$, and radius, $r_{\rm c} = (2/3^{3/2}) G M_{\rm c}/\sigma_V^2$ \citep{pflamm2009}.  The gravitational potential of the SMBH is given by
\begin{equation}
\Phi_{\rm bh} = {G M_{\rm bh} \over r}
\end{equation}
and is modeled by a sink term.  Here, we assume the gravitational potential is static - $M_{\rm bh}$ does not grow - a valid approximation since the mass accreted during the simulation is small.

We use inflow boundaries to simulate the NSC's motion through the central galaxy medium.  In order to accurately resolve the mass accretion rate on the black hole SMBH sink sizes are taken to be within hundredths of sonic radii, $r_{\rm s}$, where
\begin{equation}
r_{\rm s} = \frac{2 G M_{\rm bh}}{c_\infty^2 ( 1 + \mu_\infty^2)},
\end{equation}
 following the prescriptions of \cite{ruffert1994a,ruffert1994b}.
Models are run from an initially uniform background until a steady density enhancement forms within the NSC, which usually takes  tens of  core sound crossing times.  
Convergence tests with 
higher refinement and longer run times produce models which show similar central densities and mass accretion rates to those depicted here.  
We compute both models with adiabatic  ($\gamma = 5/3$) and  nearly isothermal ($\gamma = 1.1$) equation of states in order to test  the effects of cooling.

\begin{figure*}
\centering\includegraphics[width=0.99\textwidth]{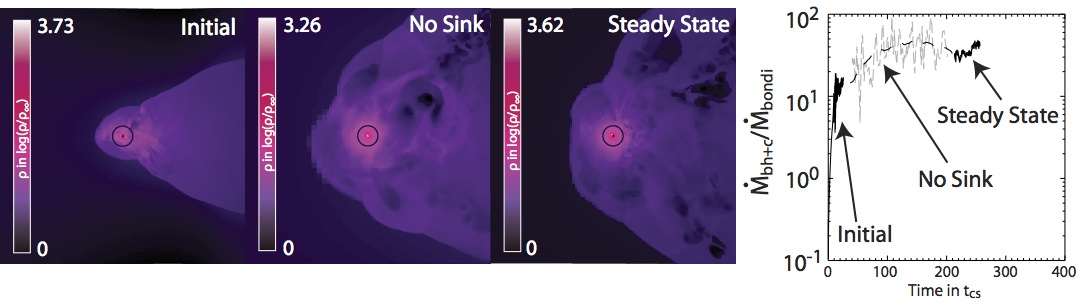}
\caption{Density contours of the flow pattern around a SMBH~+~NSC system moving through a uniform density, near isothermal ($\gamma = 1.1$) medium for three different simulation setups.  The  {\it initial}  simulation is a small scale calculation that resolves how the gas flow begins to accumulate within the NSC core and accretes  onto the fully resolved central sink.   The {\it no sink} simulation is a low resolution, large scale calculation without a sink that captures the gas build up in the NSC core  until a steady state is achieved.  The {\it steady state} simulation is a large scale calculation that includes  an embedded sink once a steady state central density enhancement has been realized. The right most line plot shows the mass accretion rate for the three different simulation setups  as a function of the sink's sound crossing time: $t_{\rm cs} = r_{\rm sink}/c_{\infty}$.  Common to all calculations  are $M_{\rm c}/M_{\rm bh} = 10$ with $M_{\rm bh} = 2.5 \times 10^7 \, M_\odot$, $\mu = 1.64$, $c_{\rm s} = 83 \, \rm{km/s}$ and $\sigma_V/c_{\rm s} = 2$.  The Plummer core radius is denoted by a black open circle and the sink size $r_{\rm sink} \approx 0.05r_{\rm b,R}$ is depicted by a black filled circle.  Here, the sink size is fractions of the modified Bondi radius as defined by  \cite{ruffert1994a}, $R_{\rm b, R} = G M_{\rm bh}/c_\infty^2$, to ensure a converged mass accretion rate onto the central SMBH \citep{ruffert1994a,ruffert1994b}.  Snapshots from left to right correspond to times $t_{\rm cs} = $~130, 221, and 250.  This simulation is identified as {\it 1A} in Figure \ref{fig:msigma} and Table \ref{table:sims}.}
\label{fig:simsExplained}
\end{figure*}

To adequately resolve the small scale sink, the significantly more extended NSC's core, and the even larger scale flow structures that develop around the cluster, a  sizable level of refinement is required on the AMR grid (commonly  14 levels of refinement).  As the minimal time step in our simulation is determined by the gas flow on small scales, necessary runtimes would need to be prohibitively large in order to simulate a resolved sink and the large scale gas structure until a steady state density enhancement forms within the NSC.
  Instead, we construct the accretion rates onto our model SMBH+NSC systems from a set of three simulation setups, as depicted in Figure \ref{fig:simsExplained} with parameters summarized in the first row of Table \ref{table:sims}.  
  
Figure \ref{fig:simsExplained} depicts  the flow of nearly isothermal gas ($\gamma = 1.1$) in the vicinity of  a SMBH ($M_{\rm bh} = 2.5 \times 10^7 \, M_\odot$, $r_{\rm sink} = 0.5$~pc) surrounded by a massive NSC ($M_{\rm c}/M_{\rm bh} = 10$, $r_{\rm c} = 5.3$~pc, $\sigma_V =  176\, \rm{km/s}$).  The gas has a sound speed of $c_\infty = 83 \, {\rm km/s}$ and the SMBH~+~NSC complex  is moving at a Mach number of $1.64$ with respect to the ambient  gas.  The ratio of $M_{\rm c}/M_{\rm bh} = 10$ is consistent with observations:  $0.1 \lesssim M_{\rm c}/M_{\rm bh}\lesssim 100$ \citep{graham2009,neumayer2012b}. However, as noted in Section \ref{section:anal},  as the ratio of $M_{\rm c}/M_{\rm bh}$ decreases so does   the ability of a NSC to significantly alter the flow of gas onto the embedded SMBH (for a given $\sigma_{V}$).  Thus, the  model  depicted in Figure \ref{fig:simsExplained} provides  an example in which the presence of a massive NSC can drastically  alter the flow properties  around the central SMBH.  

The first of these simulations, labeled {\it initial} in the density contours and mass accretion rate plots shown in  Figure \ref{fig:simsExplained}, is a small scale  simulation which follows the gas flow as  the central density enhancement begins to form in the NSC core while simultaneously tracking the mass accretion rate onto the fully resolved central sink.  Once the 
large scale bowshock begins to interact with the boundaries of the computational domain (at about three to five  core sound crossing times) the simulation is halted.   We concurrently simulate the same initial setup in a larger box (about fifty core radii) at a lower resolution and follow the  density build up within the NSC's core without the presence of a sink. We label this simulation as {\it no sink} in Figure \ref{fig:simsExplained}.  Because the presence of a sink has a minimal effect on the  build up of mass in the core, the 
central density in the  large scale simulation, albeit lower resolution, agrees well  with central  density evolution observed in the {\it initial}  simulation.   While there is no explicit sink in these second set of simulations, the mass accretion rate onto a central black hole can be relatively  accurately  inferred  from the  gas properties within the NSC's core as shown in Figure \ref{fig:simsExplained}.  Once a steady state density enhancement has formed in the central regions of the NSC, these large scale simulations are then refined further until a  central sink is resolved by at least 16 cells.  This level of refinement was chosen to allow the unsteady mass accretion rate  to converge, as argued by \cite{ruffert1994b}.   We refer to these simulations as  {\it steady state} in Figure \ref{fig:simsExplained}.  By using the set of three simulations discussed here for each SMBH+NSC system, we can resolve both the larger scale flow around the nuclear star cluster and the small scale flows into the accreting SMBH.
Because we are not explicitly including radiative cooling, the simulations depicted in Figure \ref{fig:simsExplained}  can be  easily rescaled to consider a wide range of NSC properties.  For example,  for a fixed $M_{\rm c}/M_{\rm bh}$, the structure of the flow will remain unchanged provided the ratio $\sigma_V^2/(c_{\rm s}^2 + v^2)$ remains constant \citep{naiman2011}.  

\begin{figure*}
\centering\includegraphics[width=0.99\textwidth]{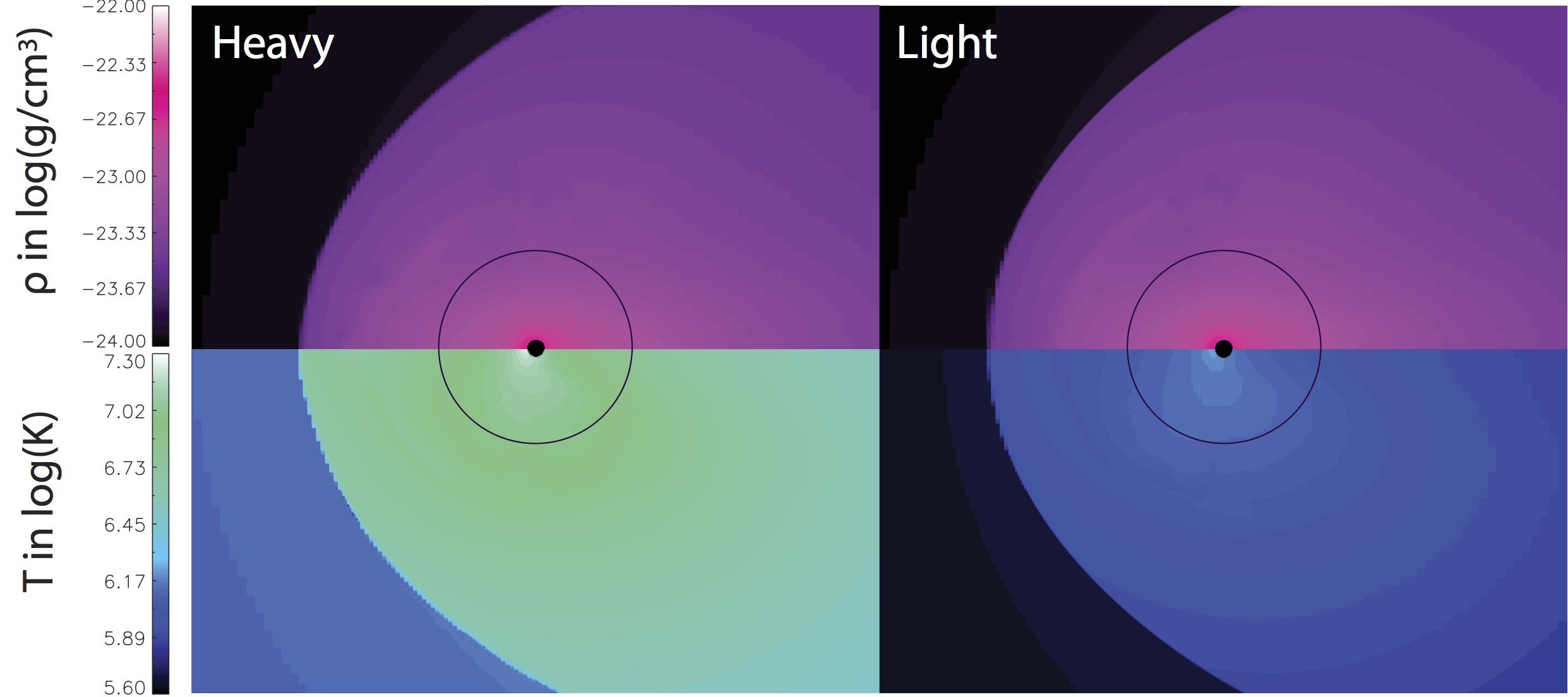}
\caption{Density (top panels) and temperature (bottom panels) contours of the flow pattern around a {\it Heavy} ({\it 2A} in Figure \ref{fig:msigma}, Table \ref{table:sims}) ($M_{\rm c} = 2.5 \times 10^8 \, M_\odot$, $M_{\rm bh} = 2.5 \times 10^7 \, M_\odot$, $\sigma_V = 280 \, {\rm km/s}$) and {\it Light} ({\it 2B}) ($M_{\rm c} = 10^8 \, M_\odot$, $M_{\rm bh} = 10^7 \, M_\odot$, $\sigma_V = 177 \, {\rm km/s}$) SMBH~+~NSC complexes moving through a uniform density in an adiabatic ($\gamma = 5/3$) medium with $\mu = 1.33$ and $\sigma_V/c_\infty = 2.12$.  The Plummer core radius is denoted by a black open circle and the sink size $r_{\rm sink} \approx 0.05r_{\rm b,R}$ is denoted by a black filled circle.}
\label{fig:scaled}
\end{figure*}

In Figure \ref{fig:scaled} we compare the flow in and around a {\it heavy} SMBH~+~NSC complex ($M_{\rm bh} = 2.5 \times 10^7 \, M_\odot$, $M_{\rm c} = 2.5 \times 10^8 \, M_\odot$, $\sigma_V = 280 \, \rm{km/s}$) moving through hot gas ($c_\infty = 140 \, {\rm km/s}$, $\mu = 1.33$, $\gamma = 5/3$) with the flow surrounding  a {\it lighter}  system ($M_{\rm bh} = 10^7 \, M_\odot$, $M_{\rm c} = 10^8 \, M_\odot$, $\sigma = 177 \, {\rm km/s}$) moving through cold gas ($c_\infty = 89 \, {\rm km/s}$, $\mu = 1.33$, $\gamma = 5/3$).  The parameters of the less massive model are chosen such that the ratio $\sigma_V^2/(c_\infty^2 + v^2)$ remains the same.   Because this ratio is constant, the flow is nearly identical.  It is important to note that while the mass accretion rate onto the central SMBH is higher for the more massive system, the accretion rate enhancement with respect to that of a black hole without the NSC, $\dot{M}_{\rm bh+c}/\dot{M}_{\rm bondi}$, is  the same between the two simulations. 

The set of simulations shown in Figure \ref{fig:simsExplained}  (Figure~\ref{fig:scaled}) depict the gas flow around a compact NSC slowly transversing  through nearly isothermal (adiabatic)  gas.  However, the enhancement in the black hole's accretion rate depends not only on  the  thermodynamical conditions of the ambient  gas but also on the properties of the NSC.  The remainder of this paper is thus devoted to calculating the necessary  conditions for large central density enhancements  in the centers of SMBH~+~NSC systems and  determining whether or not these conditions persist during a galaxy merger. 

\section{Necessary Conditions for Accretion Rate Enhancements}\label{sec:cond}

In order for the gravitational potential of the NSC to alter 
the local gas flow before it is accreted onto the central SMBH, 
the NSC must be moving relatively slowly through cold gas,  
a condition shown in \citet{naiman2011}
to be equivalent to requiring that 
\begin{equation}
\sigma_{V}^2 > c_{\rm s}^2 + v^2.
\end{equation}

\begin{figure*}
\centering\includegraphics[width=0.99\textwidth]{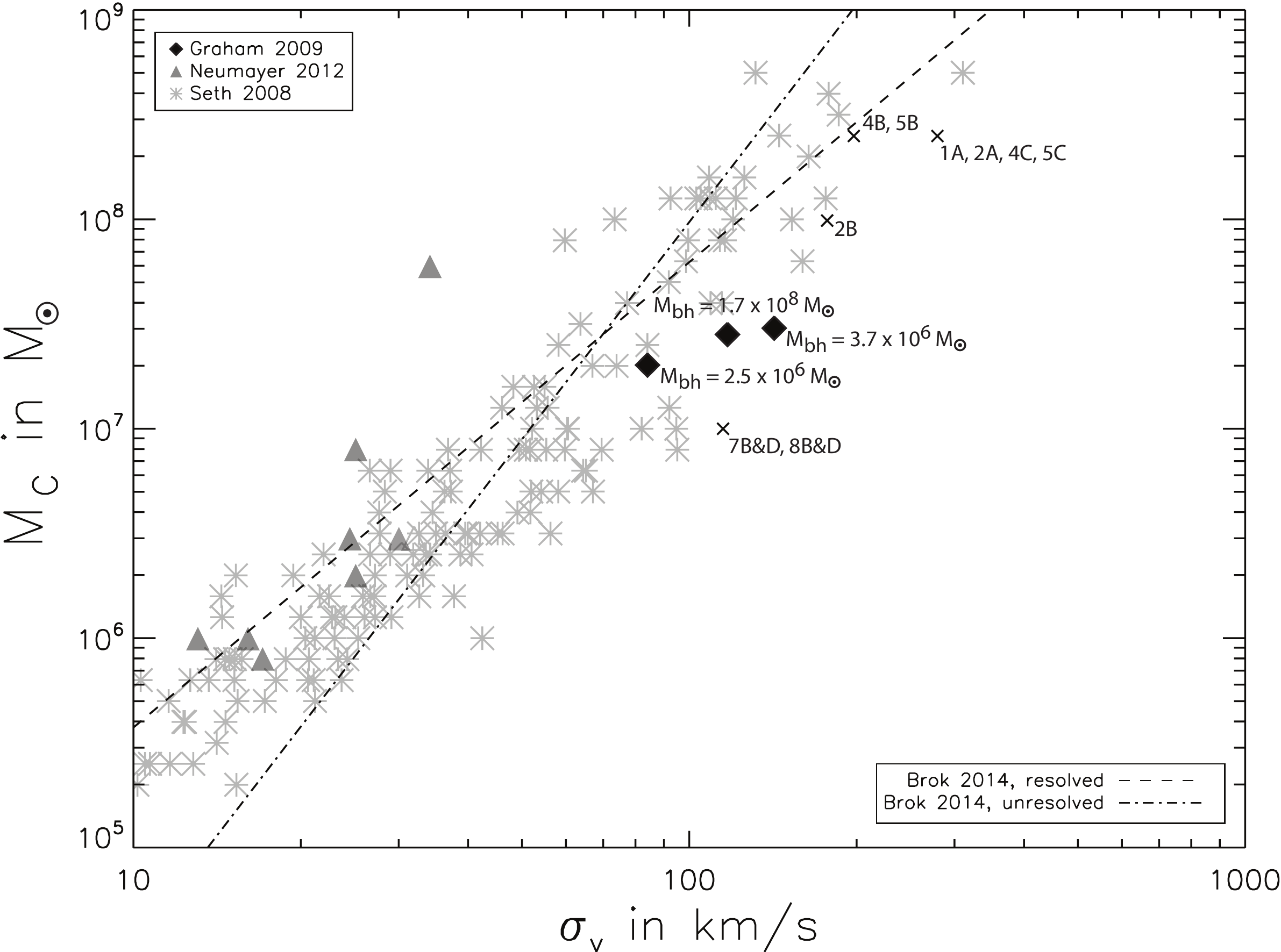}
\caption{Mass--size  relation for  NSCs.  Grey asterisks are  from \cite{seth2008} while  grey triangles are taken from the \cite{neumayer2012b} sample of NSCs in bulgeless galaxies.  The three black diamonds show measurements of NSC masses and sizes where the mass of the embedded SMBH is also known \citep{graham2009}.  The two lines show the mass-size relation inferred from the NSC sample of \cite{brok2014} with ({\it dashed} line) and without ({\it dot-dashed} line) the unresolved sources included in their fit.   Here, all half light radii have been scaled to a core radius assuming the NSC potentials are well described by a Plummer model. The mass and $\sigma_V$ of the clusters presented in the simulations of Figures \ref{fig:scaled}-\ref{fig:subgrid} are shown with black crosses.}
\label{fig:msigma}
\end{figure*}

For this reason, knowledge of the velocity dispersion of typical NSCs is vital in determining whether or not central gas densities will form in their cores (provided that  $M_{\rm c}/M_{\rm bh}\gg 1$).   Surveys of NSCs show half light radii in the range $1-5$~pc \citep{boker2010,georgiev2014}, which assuming a Plummer model (with $M/L=1$), gives a range in core radii of $0.8-3.8$~pc.  These results combined with the observed mass range of approximately $10^5 - 10^7 \, M_\odot$ \citep{boker2010,georgiev2014,neumayer2012b} naturally  result in a large range of possible velocity dispersions: $\sigma_V \approx 5 - 100 \, {\rm km/s}$.  Intuitively, we expect more massive systems to have larger core radii, a fact which is 
born out in several surveys of NSC properties as depicted in Figure \ref{fig:msigma} \citep{seth2008,brok2014}.  
If we narrow our sample size 
 by restricting our calculations to systems observed to contain both a NSC and a SMBH the mass-size relation of NSCs is less well constrained as clearly depicted in Figure \ref{fig:msigma}.
Measuring the mass of theblack hole and the mass and velocity dispersion of the NSC requires high  resolution observations, which   are only available for a handful   of systems, with the best currently known example being  our own Milky Way.
In the Galaxy's nuclei, the black hole mass has been measured to be  $M_{\rm bh} \approx 4 \times 10^6 \, M_\odot$ \citep{ghez2008,gillessen2009,genzel2010,do2013}. With a NSC mass of $M_{\rm c} \approx 3 \times 10^7 \, M_\odot$ and a half light radius of $r_{\rm eff} \approx 4$~pc, one finds $r_{\rm c} \approx 3$~pc and $\sigma_V \approx 130 \, {\rm km/s}$ \citep{graham2009,feldmeier2014}.  

The observed NSC velocity dispersions in galaxies containing black holes, which are in the range $\sigma_V \approx 100 - 300 \, {\rm km \, s^{-1}}$, are similar in magnitude to both the sound speed of the surrounding gas and 
the relative gas velocities of SMBHs  during  galaxy mergers \citep{debuhr2011,debuhr2012}.  As a result, the conditions necessary 
for efficient mass accumulation are commonly satisfied, an assertion that we will quantify below by making use of detailed galaxy merger simulations.

\begin{figure*}
\centering\includegraphics[width=0.99\textwidth]{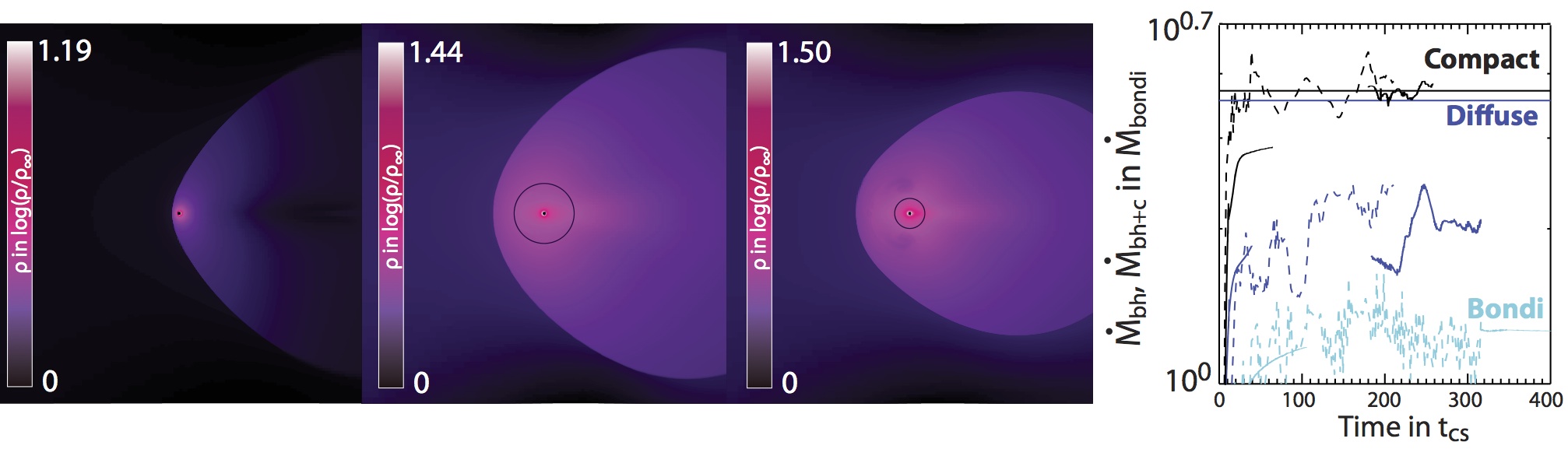}
\caption{Density contours of the flow pattern around a SMBH with $M_{\rm bh} = 2.5 \times 10^7 \, M_\odot$ moving through a uniform density medium characterized by  $\gamma = 5/3$. Simulation snapshots are plotted  for a  naked SMBH ({\it left}, {\it 4A} in Table \ref{table:sims}), a SMBH  embedded in a diffuse NSC  ({\it middle}, {\it 4B} in Figure \ref{fig:msigma}, Table \ref{table:sims}) and a SMBH  embedded in a compact  NSC  ({\it right, 4C}) together with the mass accretion rate history  in each system, which is calculated using the  three different simulation setups discussed in Figure~\ref{fig:simsExplained}. The effect of the NSC's velocity dispersion ($M_{\rm c} = 2.5 \times 10^8 \, M_\odot$) can be seen by comparing the gas flow between the {\it compact}  ($r_{\rm c} = $~5.3~pc)  and  {\it diffuse}  ($r_{\rm c} = $~10.6~pc) clusters.  The  {\it dark blue} and {\it black} horizontal lines show our  modified Bondi-Hoyle-Lyttleton (BHL) prescription  for  the mass accretion rate in the 
presence of a NSC.   All sink sizes are $r_{\rm sink} = $~0.5~pc. Here, the sound speed and Mach number are $c_\infty = 132 \, {\rm km/s}$ and $\mu_\infty = 1.33$, respectively.  The snapshots from left to right are at times $t_{\rm cs} = $~112, 108, and 81. }
\label{fig:modBondiadia}
\end{figure*}

Figure \ref{fig:modBondiadia} shows how the  accretion rate onto a SMBH is modified by the presence of  a NSC  satisfying $\sigma_V \gtrsim c_{\rm s}^2 + v^2$.   
In all calculations, the flow is assumed to behave  adiabatically. 
Without the presence of a NSC, the bowshock penetrates close to the sink boundary.  However, in the presence of a NSC, the bowshock forms at the outer boundary of the cluster's core. As discussed in \citet{lin2007} and  \cite{naiman2011}, this effectively mitigates the effects of the gas motion on the central sink and the accretion proceeds as a nearly radial inward flow. This added protection results in a moderate mass accretion rate enhancement onto the central sink, which are slightly larger for a more compact NSC.  As the flow accumulates in the NSC's potential, a  quasi-hydrostatic envelope builds up around the central sink, whose central density increases with the compactness of the NSC.  For an adiabatic flow, this density enhancement  is accompanied by an increase in the sound speed of the flow such that  $\dot{M}_{\rm nsc} \propto \rho_{\rm nsc}/c_{\rm s,nsc}^3 \approx \rho_\infty/c_\infty^3$. As a result,  the accretion rate for a stationary sink is not expected to be aided by the increase in central density in an adiabatic flow. The moderate increase in mass accretion rates for an adiabatic flow, produced mainly by changes in the flow structure,  is shown by both our analytical (see equations \ref{eq:mdot} and \ref{eq:rho})  and simulation results.

\begin{figure*}
\centering\includegraphics[width=0.99\textwidth]{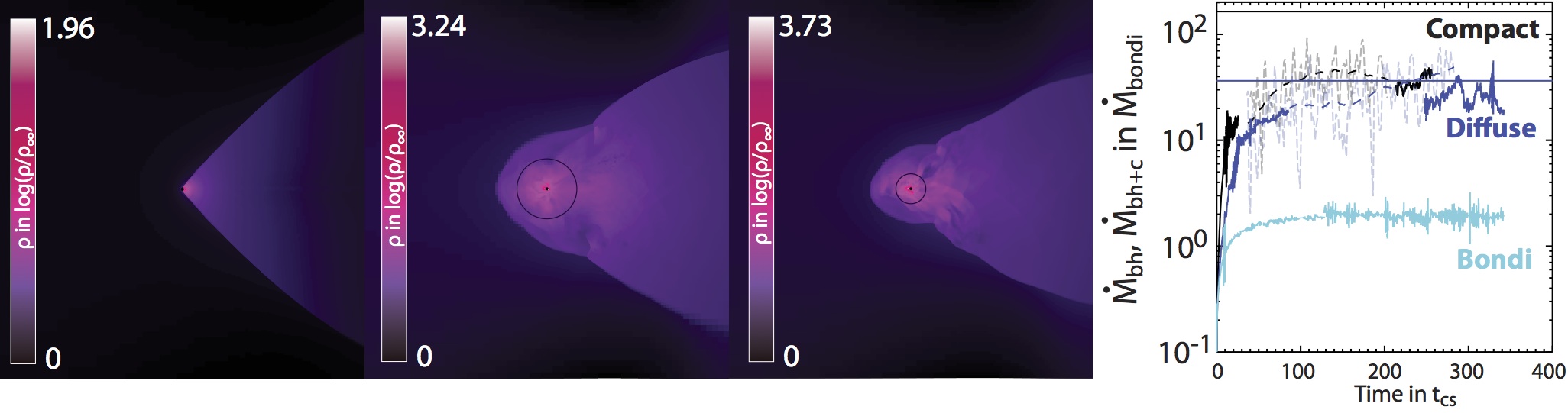}
\caption{Similar to Figure~\ref{fig:modBondiadia}  but for a near isothermal ($\gamma = 1.1$) medium. The snapshots from left to right are  at times $t_{\rm cs} = $~130, ({\it 5A}), 134 ({\it 5B}), and 131 ({\it 5C}).}
\label{fig:modBondiiso}
\end{figure*}

On the other hand, the central density enhancement enabled by the presence of a NSC  when $\sigma_V \approx c_{\rm s}^2 + v^2$ can be accompanied by a drastic increase in mass accretion rate if the gas is permitted to cool. This allows  the  central density to grow without a mitigating increase in  the local sound speed of the gas. In the near isothermal ($\gamma=1.1$) calculations depicted in  Figure \ref{fig:modBondiiso}, we have  $c_{\rm s} (r) \approx c_\infty$ and the presence of a NSC  results in an enhancement of about an order of magnitude in the mass accretion rate onto the central sink.  Here, the accretion rate fluctuates around a mean value, as the isothermal gas can collapse to much smaller scale structures than in the adiabatic case, providing the central sink with much larger  temporal changes  in the amount of accreted gas. 

Figures \ref{fig:modBondiadia} and \ref{fig:modBondiiso} together demonstrate the important effects that both the equation of state and the compactness of the NSC can have  on the  mass accreted by the central SMBH.  In what follows, we make use of cosmological simulations to estimate the range of gas and NSC properties  conducive to large enhancements in the mass accretion rate onto the central SMBHs during galaxy mergers.

\section{The Accretion History of SMBH\lowercase{s} in Galaxy  Mergers} \label{section:mergerMods}

\begin{figure*}
\centering\includegraphics[width=0.99\textwidth]{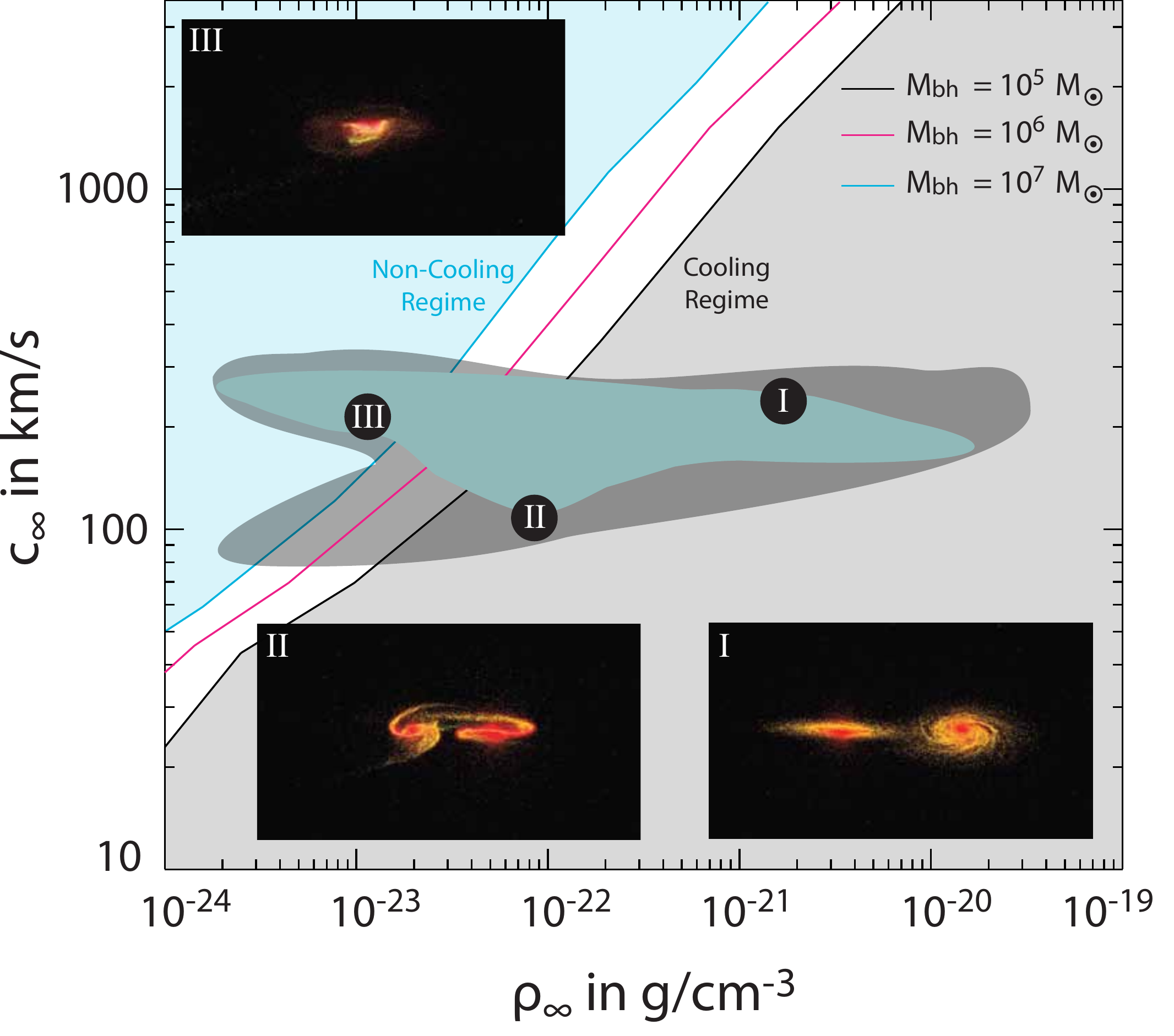}
\caption{The range of sound speeds, $c_\infty$, and densities, $\rho_\infty$, from the gas surrounding a naked SMBH in the
 galaxy merger models of \cite{debuhr2011}.  The {\it gray} contour denotes the combinations of [$c_\infty$,$\rho_\infty$]  for 
their entire  suite of models while the {\it blue} contour shows the range for their {\it fidNof} model.  The locations in the [$c_\infty$,$\rho_\infty$]  plane for three  different simulation snapshots ({\it \RNum{1}}, {\it \RNum{2}}, {\it \RNum{3}})  taken from the {\it fidNof} model of  \cite{debuhr2011} are highlighted. Simulation points {\it \RNum{1}} and {\it \RNum{2}} are at early times before the galaxies' first pass ($t_{\RNum{1}} = 0.27$~Gyrs, $t_{\RNum{2}} = 0.5$~Gyrs) and point {\it \RNum{3}} occurs right before  the merger ($t_{\RNum{3}} = 1.49$~Gyrs). The reader is referred to  Figure 1 of \citet{debuhr2011} for further details.
The plotted lines correspond to the condition $t_{\rm cool}=t_{\rm cs, acc}$ for three different  values of $M_{\rm bh}$.}
\label{fig:snapshots}
\end{figure*}

The  majority of galaxies ($50-70$\%) are expected to harbor nuclear star clusters \citep{neumayer2012}, and therefore large enhancements in accretion rates onto SMBHs  are possible during typical galaxy mergers when conditions are favorable (i.e. $\sigma_V^2 \approx c_{\rm s}^2 + v^2$).
If, in addition, during the  merger  the  gas in the central regions cools efficiently, the increase in the mass  accreted by the SMBH can be significant.  
To determine if and when these conditions are satisfied during a merger we examine the gas properties in galaxy merger simulations from \cite{debuhr2011}. These full-scale SPH simulations of major mergers include cooling, star formation and associated feedback, and feedback from a supermassive black hole. Of interest for this work is the model {\it fidNof}, which places two galaxies on a prograde parabolic orbit with unaligned spins and a merger mass ratio of 1:1. Both galaxies in the simulation have a total mass of $1.94 \times 10^{12} M_{\sun}$ (including the dark matter halo), and have $8 \times 10^5$ particles. The Plummer equivalent gravitational force softening was $47$ pc.
Figure \ref{fig:snapshots} shows the gas properties in the central core regions of the galaxies in the {\it fidNof} model. 
This range of sound speeds and densities represent the average values of a subset of particles within the accretion radii, defined as four times the simulation's gravitational softening length: $R_{\rm acc} \approx 188$~pc \citep{debuhr2011,debuhr2012}. To estimate the average properties of the  gas ($c_\infty$, $v_\infty$, $\rho_\infty$) flowing toward the SMBH  we use only particles within  a $30^\circ$ conical region in front of the SMBH's velocity vector.   While the average gas parameters are relatively insensitive to the exact value of the opening angle of the cone, the amount of inflowing gas is slightly underestimated using this method as it ignores the material which can be accreted from behind the direction of motion of the SMBH.

Given the estimated average properties of the  gas  flow at large scales, we expect the effects  of the NSC  in  altering  the mass accretion history of the central SMBH to be most prominent when the gas is able to cool efficiently, as argued in Section~\ref{sec:cond}. In a merger simulation, the condition for efficient cooling is established  when the sound crossing time across the accretion radius is longer  than the cooling time of the gas: $t_{\rm cs,acc} \gtrsim t_{\rm cool}$.  If this condition holds, the cold gas can be significantly compressed and, as a result,  lead to a large  density enhancement  in the core of the NSC.

We can estimate the accretion timescale of the SMBH as
\begin{equation}
t_{\rm cs,acc} = {2 G M_{\rm bh} \over c_\infty^3(1+\mu_\infty^2)}
\end{equation}
where $\mu_\infty = v_\infty/c_\infty$ is the Mach number of the large scale flow. The cooling time can be written as  $t_{\rm cool} = \epsilon/[n_e n_H \Lambda(T,Z)]$, where $\epsilon$ is the internal energy of the gas, $\Lambda$ is the cooling rate of the gas at a temperature $T$ and metallicity $Z = 10^{-2} \, Z_\odot$, and $n_e$ and $n_H$ are the electron and neutral hydrogen number densities, respectively.   In galaxy merger simulations, the condition $t_{\rm cs,acc} \gtrsim t_{\rm cool}$ is generally satisfied (Figure \ref{fig:snapshots}) although  
for a particular run, the average gas properties can fluctuate between the cooling and non-cooling regimes as the merger progresses. This is illustrated  in 
the simulation snapshots {\it \RNum{1}}, {\it \RNum{2}} and {\it \RNum{3}} taken  from the {\it fidNof} model of \cite{debuhr2011} at early 
($t_{\RNum{1}}=0.27$~Gyrs, $t_{\RNum{2}}=0.5$~Gyrs) and late ($t_{\RNum{3}} = 1.49$~Gyrs) times in the merger process, which are depicted  in Figure \ref{fig:snapshots}. 
As a consequence, there may be times during  the galaxy merger when cooling rates within the accretion radius are high and the mass accretion rate can be heavily augmented by the presence of a NSC, provided that $\sigma_V \approx c_\infty$.    

\begin{figure*}
\centering\includegraphics[width=0.99\textwidth]{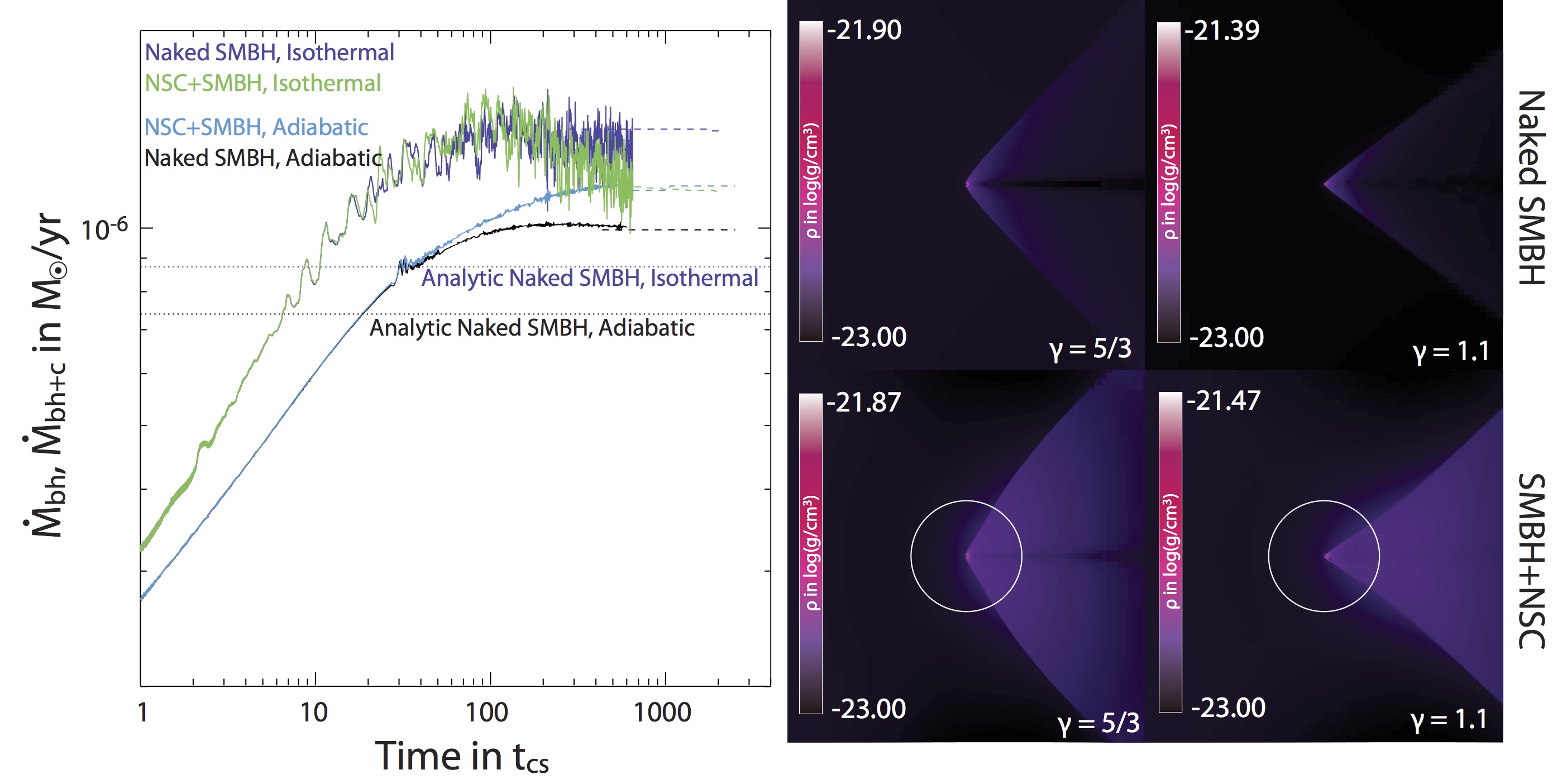}
\caption{A $M_{\rm bh} = 10^6 \, M_\odot$ black hole  with and without a surrounding NSC ($M_{\rm c} = 10^7 \, M_\odot$ and $\sigma_V=115$ km/s) propagates with $\mu = 1.5$  through a background medium with $c_{\rm s} = 200$ km/s and $\rho_\infty=10^{-23}\,{\rm g\,cm}^{-3}$ (similar to the gas properties found in simulation snapshot {\it \RNum{3}} of Figure~\ref{fig:snapshots}). The {\it left} panel  shows the mass accretion rate of models with and without a NSC for adiabatic ($\gamma = 5/3$) and near isothermal gas ($\gamma=1.1$) flows, which are calculated using different simulation setups  as discussed in Figure~\ref{fig:simsExplained}.   Under these conditions,  $\sigma_V<c_\infty$ and the gas flow around the SMBH is not altered by the presence of the NSC. As a result,  the change in mass accretion rate  between the model with and without the NSC is negligible, even when cooling is efficient.  Cluster radii are shown as white circles.  The simulations ``Naked SMBH, Isothermal", ``NSC~+~SMBH, Isothermal", ``Naked SMBH, Adiabatic" and ``NSC~+~SMBH, Adiabatic" are denoted in Table \ref{table:sims} by {\it 7A}, {\it 7B}, {\it 7C}, and {\it 7D}, respectively.  Points {\it 7B} and {\it 7D} are also shown in Figure \ref{fig:msigma}.}
\label{fig:nocool}
\end{figure*}

Since  we are not treating  the feedback from the black hole explicitly, we use both adiabatic (inefficient cooling; efficient feedback) and isothermal (efficient cooling; inefficient feedback) simulations to illustrate the effects  of the surrounding NSC on the gas flow as a whole and the importance of the $\sigma_V \approx c_\infty$ condition. 
Figure \ref{fig:nocool} shows the gas in the inner regions of a model where $\sigma_V < c_\infty$.  Here, a moderately massive black hole ($M_{\rm bh} = 10^6 \, M_\odot$) with and without a surrounding NSC ($M_{\rm c} = 10^7 \, M_\odot$ and $\sigma_v = 115 \, {\rm km/s}$) propagates  through a background medium with $c_\infty=$ 200km/s and $\rho_\infty=10^{-23}\,{\rm g\,cm^{-3}}$ (similar  to the gas properties found  in simulation  snapshot {\it \RNum{3}} of Figure \ref{fig:snapshots}).  Because $\sigma_V < c_\infty$, the gas flow around the SMBH is not altered by the presence of the NSC and  the mass accretion rates change only minimally between the model with and without the NSC, even when cooling is efficient.  This is corroborated by the results of the near isothermal and adiabatic simulations, which are  shown in Figure \ref{fig:nocool}.

\begin{figure*}
\centering\includegraphics[width=0.99\textwidth]{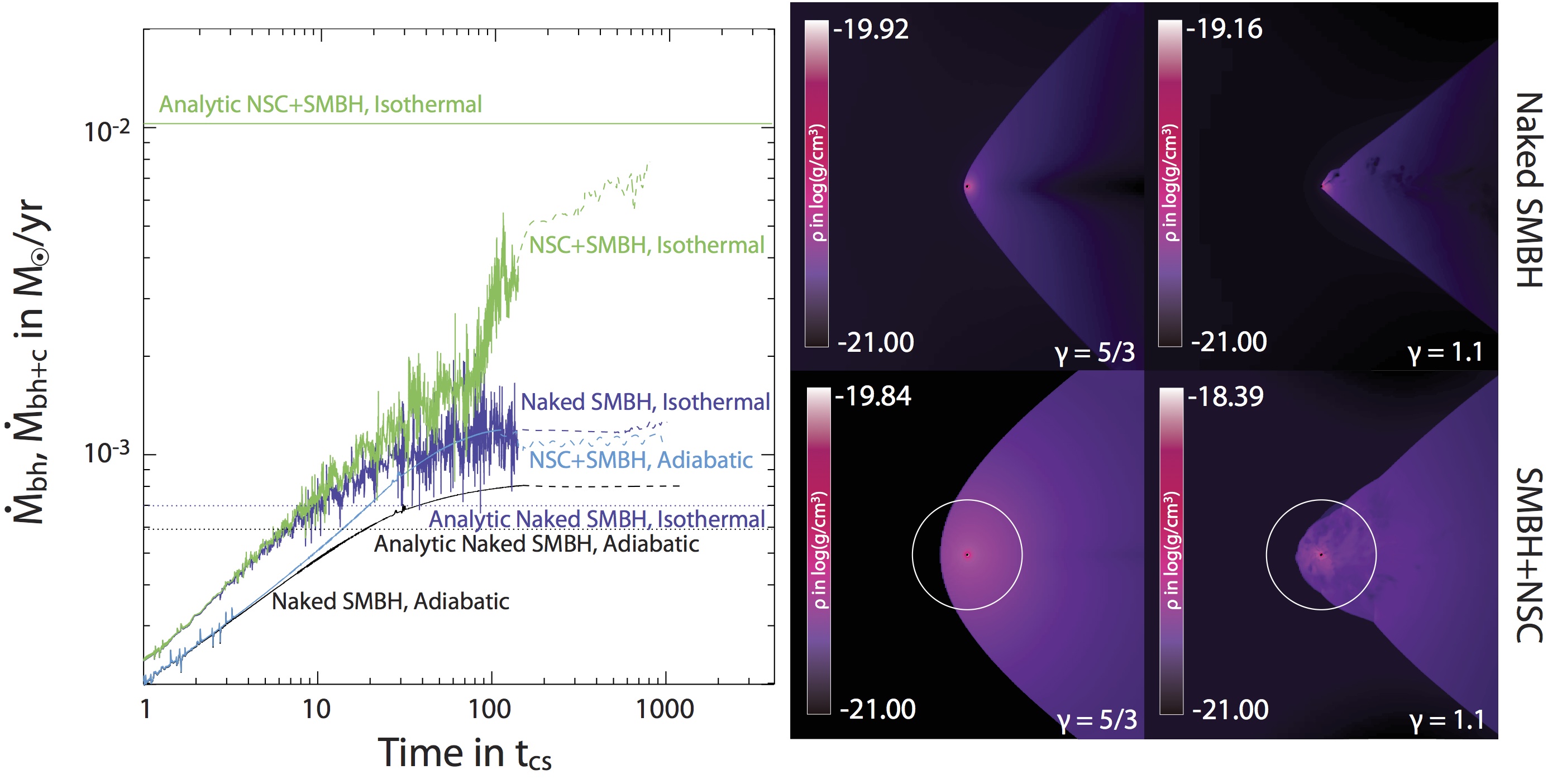}
\caption{Similar to Figure~\ref{fig:nocool} but in this case the black hole  propagates through a background medium with $c_\infty = 100$ km/s and $\rho_\infty=10^{-21}{\rm \,g\,cm}^{-3}$ (similar to those found in simulation snapshot {\it \RNum{1}} of Figure~\ref{fig:snapshots}). Because $\sigma_V>c_\infty$,  the presence of a NSC can result in a large mass feeding rate increase when compare to the case without a NSC, in particular when the gas cools efficiently ($\gamma=1.1$). Once again, the accretion rate onto the SMBH is calculated using different simulation setups as illustrated  in Figure~\ref{fig:simsExplained}. The models in this figure labeled ``Naked SMBH, Isothermal", ``NSC~+~SMBH, Isothermal", ``Naked SMBH, Adiabatic" and ``NSC~+~SMBH, Adiabatic" are denoted in Figure \ref{fig:msigma} and Table \ref{table:sims} by points {\it 8A}, {\it 8B}, {\it 8C}, and {\it 8D}, respectively.}
\label{fig:cool}
\end{figure*}

When the SMBH+NSC complex propagates into a  region in parameter space
where cooling is efficient and  the condition $\sigma_V \gtrsim c_\infty$ is satisfied,  the presence of a NSC can dramatically increase the mass supply onto the SMBH. Figure \ref{fig:cool}  shows the gas flow  in the inner regions of a NSC  where cooling is predicted to be efficient.   Similar to Figure \ref{fig:nocool}, the SMBH~+~NSC complex is characterized  by  $M_{\rm bh} = 10^6 \, M_\odot$ and $M_{\rm c} = 10^7 \, M_\odot$ ($\sigma = 115 \, {\rm km/s}$) but in this case it  propagates  through a background medium with $c_\infty=$ 100 km/s and $\rho_\infty=10^{-21}\,{\rm g\,cm^{-3}}$ (similar  to those found  in simulation snapshot {\it \RNum{1}} of Figure \ref{fig:snapshots}).  Here, even without the presence of a NSC, the accretion rate is significantly  higher than that derived from  Figure \ref{fig:nocool} due to the isothermal equation of state, higher density and lower sound speed of the medium.
However, in these efficient cooling conditions, the presence of a NSC  can have drastic effects on the  mass feeding rate enhancement  when compare to the case without a NSC. This is evident when comparing the evolution of the mass accretion rate calculated  in  the  adiabatic and near isothermal simulations.  We note here that in both regimes,  the prescription laid out by equations (\ref{eq:mdot}) and (\ref{eq:rho}) provides a relatively good estimate of the steady state mass accretion rate onto the  SMBH (see Figures \ref{fig:nocool} and  \ref{fig:cool}). In what follows, we will assume the validity of such  prescription in order to estimate the growth history of SMBHs in galaxy mergers. 

To establish the mass feeding history  of merging SMBHs embedded  in NSCs, we use  the  gas properties  in the central regions of simulated merging galaxies.  Figure \ref{fig:subgrid} shows  how the growth history of  the central SMBHs, as derived from the {\it fidNof} simulation of \cite{debuhr2011}, is  altered  by the presence of a NSC. In both galaxies, a NSC  with $M_{\rm c} = 10^7 \, M_\odot$ is assumed to reside in each galactic  center at the start of the simulation. These  values are consistent with observations of NSC around SMBHs \citep{graham2009,graham2011}.  The calculation  assumes that the flow is able to cool  efficiently  ($\gamma = 1.1$) and that the NSC is unable to grow as the mass of the SMBH increases such that the mass accretion prescription reverts to the unmodified version (equation \ref{eq:cpmmacc}) when $M_{\rm bh} > M_{\rm c}$.  At early times, the presence of the NSC enables the central SMBH to grow quicker  than it would if it was in isolation due to the increase in gas density accreted by the central SMBHs as described by equation \ref{eq:mdot}.  As the galaxy merger progresses and the mass of the SMBH quickly increases above $M_{\rm c}$, the gravitational influence of the cluster ceases to be relevant and the mass accretion prescription reverts to that used in the simulation (equation \ref{eq:cpmmacc}).  Because in our formalism  $\dot{M}_\nu \propto \Omega^{-1} \propto M^{-1/2}$, the growth rate of the pre-merger SMBH hosting a NSC is reduced  when compared to the one in isolation, whose is initially lighter.  The swifter rate of growth  for the unmodified SMBH  implies  that it is able to reach a  similar total mass  to that surrounded by a  NSC before the merger takes place.

Despite having only a brief impact, this early growth spurt induced by the presence of a NSC results in a  different early growth  and feeding history for the pre-merger SMBHs. We find that the initial properties of the NSC have an enduring effect on the luminosity and mass assembly history of the early time pre-merger SMBHs. We note here that we have assumed that the NSC's mass remains unchanged  during the entire simulation. If a larger stellar concentration is able to form around the growing SMBH, then the evolving NSC could have a longer lasting impact. 

\section{Summary and Conclusions} \label{section:summary}

An understanding of how matter can be funneled to galactic nuclei is  essential when constructing a cosmological framework for galaxy evolution. When modeling galaxy evolution, an implementation of a sub-grid mass accretion prescription to estimate the gas flow onto SMBHs is required for the sake of computational efficiency. 
In simulations of merging galaxies, some form of the classical Bondi-Hoyle-Lyttleton (BHL) accretion prescription is usually implemented to estimate the SMBH's feeding rate. This prescription assumes that the properties of gas at hundreds of parsecs  accurately determine the mass accretion rate. In this paper we argue that NSCs, a common component of galactic centers at parsec scales, can provide an efficient mechanism for funneling gas towards the SMBH at scales which are commonly unresolved in cosmological simulations.  

\begin{figure*}
\centering\includegraphics[width=0.99\textwidth]{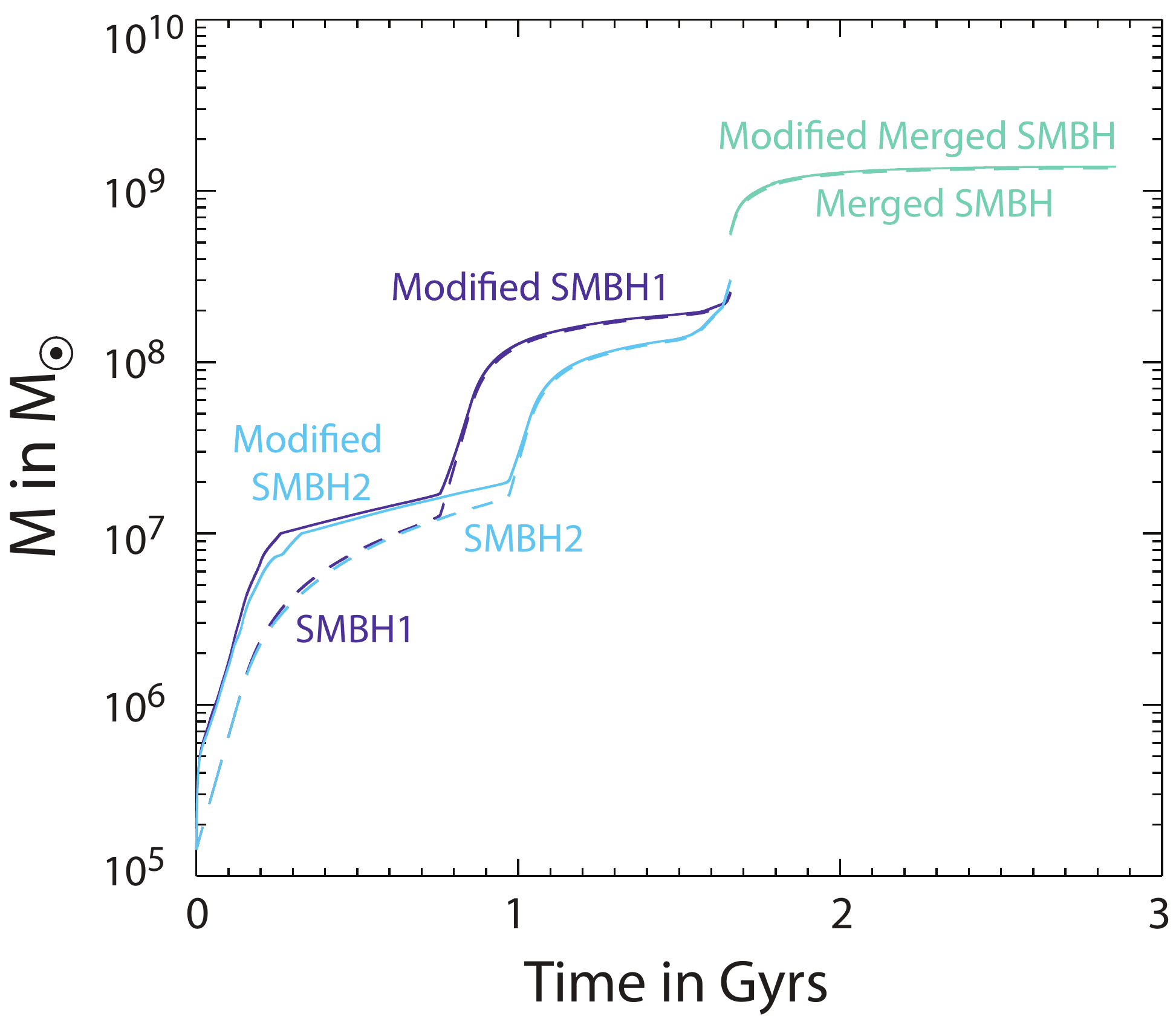}
\caption{The growth history of the two central SMBHs in the merging galaxy model {\it fidNof} of \cite{debuhr2011}. The prescription laid out by equations (\ref{eq:mdot}) and (\ref{eq:rho}) is used to estimate the  steady state mass accretion rate onto the SMBH, which in turn use  the gas properties as derived by the SPH simulations to calculate the augmented growth of the SMBH masses during the galaxy merger simulation. A NSC characterized by $M_{\rm c} = 10^7 \, M_\odot$ and $\sigma_V = 180 \, {\rm km/s}$ \citep[consistent with observations of SMBH+NSC systems]{graham2009,graham2011} is assumed to reside in each galactic center at the start of the simulation. The calculation assumes that the flow is able to cool efficiently ($\gamma= 1.1$) and that the mass of the NSC is fixed. As the galaxy merger evolves and the mass of the SMBHs increase above $M_{\rm c}$, the gravitational influence of the NSC stops being relevant.}
\label{fig:subgrid}
\end{figure*}

For the conditions expected to persist in the centers of merging galaxies, the resultant large central gas densities in NSCs should produce  enhanced accretion rates onto the embedded SMBHs, especially if cooling is efficient. 
Because these NSCs are typically more massive than the central SMBH, they can significantly alter the gas flow before being accreted.  While the model shown in Figure \ref{fig:subgrid}  results in a modest increase in the final mass of the merged SMBH, the presence of NSCs result in faster SMBH  growth rates and higher bolometric luminosities than predicted by the standard BHL formalism.  Obviously, these calculation are incomplete and would improve with a self-consistent implementation of  feedback. 

It has been suggested that the interplay between SMBH mass and host galaxy properties indicates that black hole feedback during mergers alters the properties of gas at galactic scales in order to shape these observed correlations \citep{johansson2009,debuhr2011,debuhr2012}. Progress in our understanding of these processes and higher resolution simulations  will be necessary before we can conclude that quasar feedback is in fact an essential ingredient. With more accurate simulations of the growth of SMBHs surrounded by NSC in galaxy mergers, we can better constrain the relevant physics responsible for the $M_{\rm bh}-\sigma$ and $M_{\rm bh}-L$ relations from comparisons to observational data. \\
\\
\indent We acknowledge helpful discussions with Doug Lin, Elena D'Onghia, Anil Seth, Morgan MacLeod and Lars Hernquist. We also thank the anonymous referee for constructive comments. Support was provided by the David and Lucile Packard Foundation and NSF grant AST-0847563, Simons Foundation and NASA grant NNX11AI97G. 

\clearpage

\begin{deluxetable}{ccccccccc}
\tablehead{\colhead{Name} & \colhead{$M_{\rm c}$ $[10^8 \, M_\odot]$} & \colhead{$M_{\rm bh}$ $[10^7 \, M_\odot]$} & \colhead{$\sigma_V$ $[\rm{km s^{-1}}]$} & \colhead{$r_{\rm c}$ [pc]} & \colhead{$\gamma$} & \colhead{$\mu$} & \colhead{$\sigma_V/c_\infty$} }
\startdata
1A & 2.5 & 2.5 & 280 & 5.3 & $1.1$ & $1.64$ & 1.31  \\
2A ({\it Heavy}) & 2.5 & 2.5 & 280 & 5.3 & $5/3$ & 1.33 & 2.12 \\
2B ({\it Light}) & 1.0 & 1.0 & 177 & 5.3 & $5/3$ & 1.33 & 2.12  \\
4A & 0 & 2.5 & Naked BH & Naked BH & $5/3$ & 1.33 & Naked BH  \\
4B & 2.5 & 2.5 & 198 & 10.6 & $5/3$ & 1.33 & 1.50  \\
4C & 2.5 & 2.5 & 280 & 5.3 & $5/3$ & 1.33 & 2.12  \\
5A & 0 & 2.5 & Naked BH & Naked BH & 1.1 & 1.64 & Naked BH  \\
5B & 2.5 & 2.5 & 198 & 10.6 & 1.1 & 1.64 & 0.93  \\
5C & 2.5 & 2.5 & 280 & 5.3 & $1.1$ & $1.64$ & 1.31  \\
7A & 0 & 0.1 & Naked BH & Naked BH & $1.1$ & $1.5$ & Naked BH  \\
7B & 0.1 & 0.1 & 115 & 2.0 & $1.1$ & $1.5$ & 0.58  \\
7C & 0 & 0.1 & Naked BH & Naked BH & $5/3$ & $1.5$ & Naked BH  \\
7D & 0.1 & 0.1 & 115 & 2.0 & $5/3$ & $1.5$ & 0.58  \\
8A & 0 & 0.1 & Naked BH & Naked BH & $1.1$ & $1.5$ & Naked BH  \\
8B & 0.1 & 0.1 & 115 & 2.0 & $1.1$ & $1.5$ & 1.2  \\
8C & 0 & 0.1 & Naked BH & Naked BH & $5/3$ & $1.5$ & Naked BH  \\
8D & 0.1 & 0.1 & 115 & 2.0 & $5/3$ & $1.5$ & 1.2  

\enddata
\tablecomments{Columns are (1) The name of the simulation - figure denoted by a number, subplot or line denoted by a letter, (2) mass of the NSC, (3) mass of the SMBH, (4) velocity of the NSC, (5) NSC cluster radius, (6) adiabatic index of the ambient gas, (7) Mach number of the flow, and (8) the ratio of NSC velocity dispersion to background sound speed.  Simulations with out a NSC, and thus no parameter for the NSC mass, velocity dispersion or radius, are denoted by $M_{\rm c} = 0.0$ and the words ``Naked BH" in all other fields.}
\label{table:sims}
\end{deluxetable}

\end{document}